\documentclass[onecolumn,aps,showpacs]{revtex4}
\usepackage{graphicx}
\input{epsf}
\begin{document}

\title{Very low 1/f noise at room temperature in fully epitaxial Fe/MgO/Fe magnetic tunnel junctions}

\author{F.G. Aliev, R. Guerrero, D. Herranz, R. Villar}
\affiliation{Departamento de Fisica de la Materia Condensada, C-III,
Universidad Autonoma de Madrid, 28049, Madrid, Spain}

\author{F. Greullet, C. Tiusan, M. Hehn}
\affiliation{Laboratoire de Physique des Mat\'eriaux, UMR CNRS 7556,
Nancy Universit\'e, Bd. des Aiguillettes, B.P. 239, 54506 Vand{\oe
}uvre-l\`es-Nancy Cedex, France}

\date{\today}

\begin{abstract}

We report on room temperature 1/f noise in fully epitaxial
Fe(45nm)/MgO(2.6nm)/Fe(10nm) magnetic tunnel junctions (MTJs) with
and without carbon doping of the Fe/MgO bottom interface. We have
found that the normalized noise (Hooge factor) asymmetry between
parallel and antiparallel states may strongly depend on the applied
bias and its polarity. Both types of MTJs exhibit record low Hooge
factors being at least one order of magnitude smaller than
previously reported.

\end{abstract}

\maketitle

Recent advances in spintronics are mainly related to progress in the
understanding of tunneling magnetoresistance (TMR) phenomena
\cite{Julliere75,Moodera95,Miyazaki95}. The major step forward was
taken a few years ago when, following predictions by the
theory\cite{Butler01,Mathon01}, experimentalists
demonstrated\cite{Bowen01,FVincent03,Parkin04,Yuasa04,Tiusan06}
coherent spin-dependent tunneling in single crystalline
Fe(100)/MgO(100)/Fe(100) magnetic tunnel junctions (MTJs). This
achievement opened ways to create devices with extremely high room
temperature (RT) TMR exceeding 100\%. The low frequency noise at RT,
and especially the so-called 1/f noise, determines the values of the
signal to noise ratio for each of the magnetic states (parallel (P)
vs. antiparallel (AP)) of the junctions, and therefore is an
important parameter to be optimized. It was recently found that in
MTJs the 1/f noise in P and AP configurations is mainly due to
defects inside the insulating barrier\cite{Nowak99}. Previous
studies of the 1/f noise in MTJs with a MgO(111) barrier revealed a
dependence of 1/f noise on the magnetic alignment \cite{Guerrero05}.
Similar effects were later reported for sputtered grown
CoFeB/MgO(100)/CoFeB MTJs \cite{JAP06,APL07} indicating that the low
frequency noise in MTJs with a MgO barrier may not just be
determined by the derivative of tunneling resistance with respect to
the magnetic field \cite{APL07_Mazumadar}. The Fe/MgO/Fe MTJs are
ideal devices to engineer the chemical and electronic structure of
the Fe/MgO interface in order to manipulate the voltage variation of
the TMR in amplitude and sign\cite{Tiusan06}.

This letter presents a \emph{detailed study of 1/f noise in fully
epitaxial MTJs}with large RT-TMR (above 100\%). We have investigated
the normalized 1/f noise (Hooge factor $\alpha $) in Fe/MgO/Fe MTJs,
with and without carbon doping of the Fe/MgO interface, as a
function of magnetic configuration and applied bias up to 0.5V. For
carbon-doped MTJs the Hooge factor asymmetry, $\alpha
$(AP)/$\alpha$(P), may strongly depend not only on the bias but also
on its polarity. \emph{The normalized 1/f noise for both types of
the epitaxial MTJs shows record low values.}

Our epitaxial Fe(45nm)/ MgO(2.6nm)/ Fe(10nm)/ Co(20nm)/ Pd(10nm)/
Au(10nm) samples were grown by molecular beam epitaxy on MgO(100)
substrates under UHV conditions ($4.10^{-11}$ mbar base pressure). A
complete review of the growth procedure can be found in
Ref.\cite{Tiusan07}. Two different sets of the samples were grown :
with clean Fe/MgO bottom interfaces (B-MTJs) and with carbon doping
(A-MTJs) at bottom Fe/MgO interface (Fe/Fe-C/MgO). The samples with
carbon at the bottom Fe(001) electrode present an additional
$c(2\times$2) surface reconstruction\cite{Tiusan07}. The Reflection
High Energy Electron Diffraction (RHEED) patterns showed no clear
evidence of any structural difference between the two systems. After
the growth of the multilayer stack, MTJs with micrometric lateral
size were patterned using standard optical lithography and ion
etching processes.

Tunneling magnetoresistance, dynamic conductance $G(V)$ and low
frequency noise vs. bias and magnetic field have been studied at
300K using a four-probe method. The MTJs were biased using a
constant \textit{DC} current with a superimposed low amplitude
square wave ($V_{AC}<10\mu$A). The positive bias corresponds to
electrons injected from the bottom to the top electrode. The voltage
drop on the junctions and the current were obtained by using an
analogue-digital converter (ADC) which provides the dynamic
conductance and the \textit{DC} voltage. The noise measurements were
performed using a cross-correlation technique. More details of the
experimental setup were published in Refs.
\cite{Guerrero05,Guerrero06}.


Figure 1 presents a typical voltage noise spectrum $S_{V}(f)$ below
1kHz in P and AP configurations for one of the MTJs with a low noise
level. The white noise observed above 100Hz in the P state is well
accounted by the thermal and shot noise contributions. The low
frequency part of the noise spectrum is clearly dominated by the
so-called 1/f noise (see the line drawn as a guide for the eyes). In
the AP state, in the frequency range studied, the voltage noise
typically consists of the 1/f background superimposed by the
additional noise contribution of Lorentzian-type. In the following,
we shall analyze the 1/f contribution as normalized noise power by
means of the widely used phenomenological Hooge parameter ($\alpha$)
defined as $\alpha =fAS_{V}(f)/V^{2}$, where $A$ is the junction
area and V is the DC voltage applied to the junction \cite{kogan}.
Figure 1b shows the typical dependence of the Hooge parameter on the
magnetic field in A-MTJs for positive bias. A clearly visible excess
of the normalized noise in the AP state is observed for this bias
polarity. A similar feature has been found in other MTJs with MgO
barrier\cite{Guerrero05,JAP06} and has no direct link to the
derivative of the resistance in respect to the magnetic field. This
means that the noise variation between the P and AP states is not
due to magnetic domains or other magnetic inhomogeneities. We have
observed that the noise asymmetry is independent on bias polarity in
the B-MTJs. However, in the A-MTJs, where the TMR may be inverted
for one of the bias polarities, an interesting behavior of the noise
asymmetry vs. bias with normalized noise in the P state exceeding
the one in the AP state (Fig. 1c) has been measured for negative
voltages exceeding -300mV.

Most of the B-MTJs showed an increase of the normalized voltage
noise for applied biases exceeding 200mV (not shown). For some of
the B-MTJs the noise enhancement with increasing bias could even
become irreversible. The increase of the voltage noise was a huge
(up to few orders of magnitude) once bias above 200mV had been
applied, and was accompanied by some reduction (down to 120\%) in
the TMR. We explain this by the strong sensitivity of the defects
distribution in the B-MTJs to an applied electric field exceeding
10$^{6}$ V/cm. The A-MTJs showed, however, the normalized 1/f noise
decreasing with increasing applied voltage up to at least 500mV
(Fig.2a). This dependence, as is evidenced in Figures 1b and 1c, is
rather asymmetric being the inversion of the noise asymmetry linked
to the TMR inversion point (see vertical dotted line in Fig.2.) We
note that in general for the A-MTJs, both conductance and 1/f noise
level appeared to be much more stable to the applied external
electric field above 10$^{6}$ V/cm, in agreement with our recent
findings for carbon-doped epitaxial Fe/MgO/Fe with large MgO
barriers\cite{GuerreroAPL07}.


It may be seen that the normalized noise dispersion in the AP state
decreases rapidly when applying negative biases of a magnitude above
the TMR inversion (Fig. 2b), while for the P configuration the noise
dispersion is almost independent of bias. This unexpected behavior
could be a consequence of a strong enhancement of the AP conductance
once the applied bias exceeds -300mV. In our view the observed
strong suppression of the normalized "noise of the noise"
\cite{kogan} arises from a transition between the regime where the
noise in the AP conductance is mainly controlled by the defects
inside the barrier to the one in which the AP conductance and its
noise are determined by the resonant electrons tunneling to
interface localized states, contributing to opening of the new
conductance channels at negative biases exceeding -300mV.

Figure 3 resumes electron conductance (part b) and noise (part a)
data obtained for 13 MTJs (seven of the type A and six of the type
B). Both types of MTJs show rather low dispersion in the TMR values,
being 157 $\pm$ 7$\%$ for B-MTJs and 152 $\pm$ 16$\%$ for A-MTJs.
Interestingly, carbon doping does not substantially influence the
resistance by area (RA) product which was observed to be 40$\pm$ 8
k$\Omega\cdot\mu$m$^2$ for B-MTJs and 38 $\pm$ 4
k$\Omega\cdot\mu$m$^2$ for A-MTJs. These values of the TMR and
conductance are rather close to those reported previously for fully
epitaxial Fe/MgO/Fe MTJs with similar MgO barrier
thickness\cite{Yuasa04,Tiusan07}.


Our most important finding is that the normalized noise for both A-
and B- MTJs studied may be notably (about one order of magnitude)
smaller than the best previously reported levels in MTJs with MgO
barriers, which were measured at similar biases and were
characterized by somewhat larger RA ratios \cite{APL07}.
Furthermore, the smallest P-state Hooge factors observed here
($\alpha$ $\approx$ 10$^{-11}\mu$m$^{2}$) are at least one order of
magnitude smaller than the smallest Hooge factors ever reported for
MTJs with different barriers and with RA products within the wide
range between $10^{-4}$ to $10^{3}$ M$\Omega\cdot\mu$m$^{2}$ (see
dotted line in Figure 3 obtained from the analysis made in
Ref.\cite{JAP06}).

The observed very low 1/f noise in fully epitaxial
Fe(45nm)/MgO(2.6nm)/Fe(10nm) MTJs with or without carbon doping of
the Fe/MgO interface seems to be related to a much higher degree of
epitaxy of the MgO barrier in comparison with quasi epitaxial MTJs
with MgO barrier studied before. In addition, the carbon doping,
through the $c(2\times$2) reconstruction\cite{Tiusan07}, could
partially relax stress at the Fe/MgO interface, remove defects out
of MgO and create the enhanced robustness of the MgO barrier to
applied electric field.

In conclusion, we have found very low values of the room temperature
1/f noise in fully epitaxial Fe/MgO/Fe magnetic tunnel junctions.
The TMR inversion observed in A-MTJs is accompanied by an inversion
of the Hooge factor asymmetry between the P and AP states. These
results show a great potential for integration of the fully
epitaxial MTJs into spintronic devices.

The authors thank T.McColgan for critical reading of the manuscript
and acknowledge support by Spanish-French Integrated Action project
(HF2006-0039), Spanish MEC(MAT2006-07196) and Comunidad de Madrid
(S-505/MAT0194). This work, as a part of the European Science
Foundation EUROCORES Programme 05-FONE-FP-010-SPINTRA, was also
supported by funds from the Spanish MEC (MAT2006-28183-E) and the EC
Sixth Framework Programme, under Contract No. ERAS-CT-2003-980409.

\newpage

\textbf{Figure Captions.}\\

Fig.1 (a) Typical noise spectrum measured in A-MTJs in the P and AP
states with a positive bias of 200mV. Field dependence of the
normalized noise ($\alpha$) for positive (b) and negative (c)
biases. The arrows show the direction of variation of the magnetic
field.\\

Fig.2 (a) Bias dependence of the Hooge parameter in the P and AP
states in the A-MTJs, evaluated for the frequency range 2-20Hz (left
axis), and TMR vs. bias (right axis). (b) Bias dependence of the
normalized dispersion of the Hooge parameter $\Delta \alpha
/\alpha=\sqrt{<(\alpha -<\alpha >)^{2}>}/<\alpha>$ in the AP and P
states of the A-MTJs evaluated outside the transition regions
between P and AP configurations.\\

Fig.3 Normalized noise measured at +200mV (a) and zero-bias TMR (b)
as a function of RA product. The dashed line (part a) indicates the
lowest Hooge factor value previously reported for MTJs with RA
product within the wide range between $10^{-4}$ to $10^{3}$
M$\Omega\cdot\mu$m$^{2}$ (see Ref.\cite{JAP06}).

\newpage

\begin{figure}[t]
\begin{center}
\includegraphics[width=8.5cm]{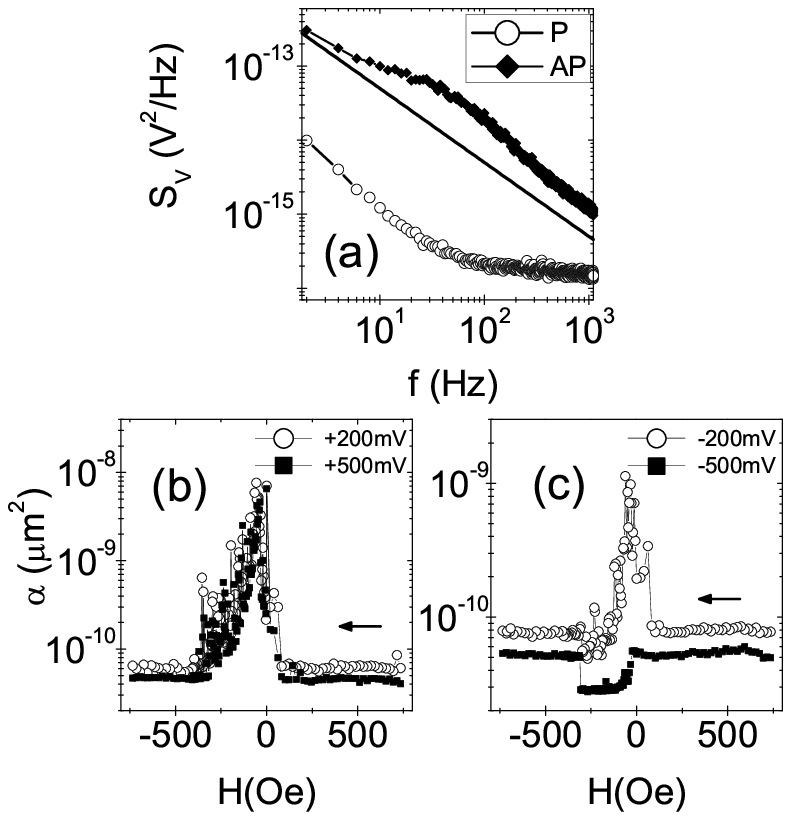}
\caption{
} \label{Fig1}
\end{center}
\end{figure}

\newpage

\begin{figure}[t]
\begin{center}
\includegraphics[width=8.5cm]{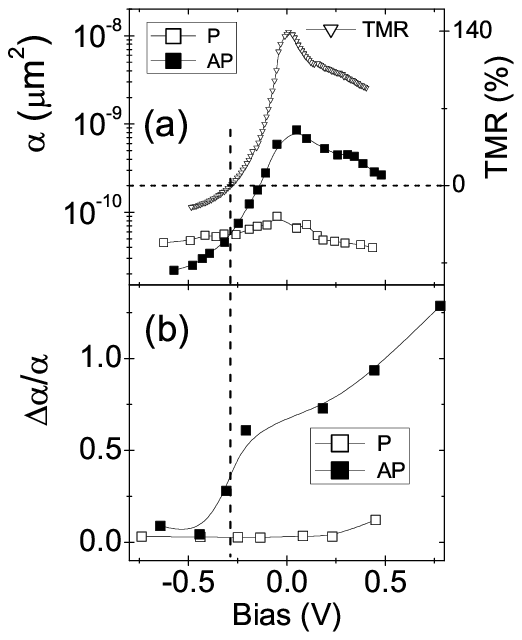}
\caption{
} \label{Fig2}
\end{center}
\end{figure}

\newpage

\begin{figure}[t]
\begin{center}
\includegraphics[width=8.5cm]{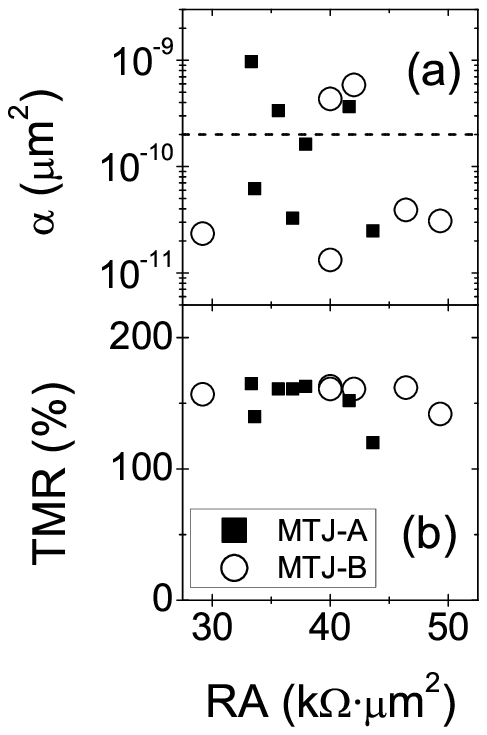}
\caption{
} \label{Fig3}
\end{center}
\end{figure}

\end{document}